\documentclass[aps,pra,reprint,superscriptaddress,amsmath,amssymb,floatfix]{revtex4-2} 

\usepackage{graphicx}
\usepackage{dcolumn}
\usepackage{bm}
\usepackage[colorlinks=true,citecolor=blue,linkcolor=blue,urlcolor=blue]{hyperref}
\usepackage[nameinlink]{cleveref}
\usepackage{leftindex}
\usepackage[utf8]{inputenc}
\usepackage[locale=UK]{siunitx}
\usepackage{booktabs}
\usepackage{prafonts}

\begin{document}

\title{Parts-per-million-accurate determination of the \texorpdfstring{$K{\alpha}$}{K-alpha} photoionization resonance of Be-like oxygen \texorpdfstring{\\}{ } with resolution of its \texorpdfstring{$^{16}$O -- $^{18}$O}{16O -- 18O} isotopic shift}

\author{Jonas Danisch}
\email{jonas.danisch@mpi-hd.mpg.de}
\affiliation{Max-Planck-Institut f\"ur Kernphysik, Saupfercheckweg 1, 69117 Heidelberg, Germany}%

 \author{Marc Botz}%
\affiliation{Max-Planck-Institut f\"ur Kernphysik, Saupfercheckweg 1, 69117 Heidelberg, Germany}%
\affiliation{Heidelberg Graduate School for Physics, Ruprecht-Karls-Universität Heidelberg, Im Neuenheimer Feld 226, 69120 Heidelberg, Germany}%

\author{Chintan Shah}%
\email{chintan.shah@mpi-hd.mpg.de}
\affiliation{NASA/Goddard Space Flight Center, 8800 Greenbelt Road, Greenbelt, MD 20771, USA}%
\affiliation{Department of Physics and Astronomy, Johns Hopkins University, Baltimore, MD 21218, USA}
\affiliation{Max-Planck-Institut f\"ur Kernphysik, Saupfercheckweg 1, 69117 Heidelberg, Germany}%

\author{Moto Togawa}
\affiliation{Max-Planck-Institut f\"ur Kernphysik, Saupfercheckweg 1, 69117 Heidelberg, Germany}%
\affiliation{European XFEL, Holzkoppel 4, 22869 Schenefeld, Germany}%

\author{Joschka Goes}
\affiliation{Max-Planck-Institut f\"ur Kernphysik, Saupfercheckweg 1, 69117 Heidelberg, Germany}%

\author{Dominic Hache}
\affiliation{Max-Planck-Institut f\"ur Kernphysik, Saupfercheckweg 1, 69117 Heidelberg, Germany}%

\author{Filipe Grilo}%
\affiliation{Laboratory of Instrumentation, Biomedical Engineering and Radiation Physics (LIBPhys-UNL), Department of Physics, NOVA School of Science and Technology, NOVA University Lisbon, 2829-516 Caparica, Portugal}%
\affiliation{Max-Planck-Institut f\"ur Kernphysik, Saupfercheckweg 1, 69117 Heidelberg, Germany}%

\author{Pedro Amaro}%
\email{pdamaro@fct.unl.pt}
\affiliation{Laboratory of Instrumentation, Biomedical Engineering and Radiation Physics (LIBPhys-UNL), Department of Physics, NOVA School of Science and Technology, NOVA University Lisbon, 2829-516 Caparica, Portugal}%

\author{{Vladimir A.} Yerokhin}%
\affiliation{Max-Planck-Institut f\"ur Kernphysik, Saupfercheckweg 1, 69117 Heidelberg, Germany}%

\author{Steffen K\"uhn}%
\affiliation{Max-Planck-Institut f\"ur Kernphysik, Saupfercheckweg 1, 69117 Heidelberg, Germany}%

\author{Awad Mohamed}%
\affiliation{%
 {Physics Division, School of Science and Technology, Università di Camerino, Via Madonna delle Carceri 9, Camerino, MC, Italy}
}%
\affiliation{%
 CNR - Istituto Struttura della Materia (ISM), 34149 Trieste, Italy
}%

\author{Roberta Totani}%
\affiliation{%
CNR - Istituto Struttura della Materia (ISM), 34149 Trieste, Italy
}%
\affiliation{%
 Elettra - Sincrotrone Trieste S.C.p.A,  Strada Statale 14, km 163,5 AREA Science Park 34149 Basovizza, Trieste, Italy
}%

\author{Monica de Simone}%
\affiliation{%
CNR - Istituto Officina dei Materiali (IOM), 34149, Trieste, Italy
}%

\author{Stefano Orlando}%
\affiliation{%
CNR - Istituto Struttura della Materia (ISM), 34149 Trieste, Italy
}%

\author{Thomas Pfeifer}
\affiliation{Max-Planck-Institut f\"ur Kernphysik, Saupfercheckweg 1, 69117 Heidelberg, Germany}%

\author{Fabrizio Nicastro}
\affiliation{Observatorio Astronomico di Roma-INAF, Via di Frascati 33, 1-00040 Monte Porzio Catone, RM, Italy}

\author{Marcello Coreno}%
\affiliation{%
CNR - Istituto Struttura della Materia (ISM), 34149 Trieste, Italy
}%
\affiliation{%
 Elettra - Sincrotrone Trieste S.C.p.A,  Strada Statale 14, km 163,5 AREA Science Park 34149 Basovizza, Trieste, Italy
}%

\author{Jos\'e R.~{Crespo L\'opez-Urrutia}}
\email{crespojr@mpi-hd.mpg.de}
\affiliation{Max-Planck-Institut f\"ur Kernphysik, Saupfercheckweg 1, 69117 Heidelberg, Germany}%

\date{\today}

\begin{abstract}
We determine with high accuracy the energy of the inner-shell transition $1s^2 2s^2~^1\mbox{S}_0 \rightarrow 1s 2s^2 2p_{3/2}~^1\mbox{P}_1$ $^{16}\mbox{O}_{K{\alpha}}^{4+}$ at 554.372(3)\,eV ($\lambda =$\,22.36480(12)\,\AA ) as well as its small shift of $2.2\pm \SI{1.3}{\milli \eV}$ ($\Delta\lambda =$\,0.089(52)\,m\AA ) for the $^{18}$O isotope. This transition blends with a $K_\alpha$ line of O$^{5+}$  used in astrophysical diagnostics, potentially affecting its reliability.  In contrast to our experimental uncertainty of $\pm$3\,meV, advanced electronic structure predictions for this four-electron system, including quantum electrodynamic (QED) corrections on the order of 100\,meV, still scatter by more than $\pm$250\,meV. 
Ions generated and stored in an electron beam ion trap were excited at the ELETTRA synchrotron facility with monochromatic soft x rays, with photon energies corrected by an additional spectrometer. Upon resonant excitation of O$^{4+}$ and their subsequent autoionization, we separate the photoions of each isotope by a time-of-flight measurement. This way, we resolve soft x-ray isotopic shifts of a few meV, obtain very accurate data on essential astrophysical ion, and test calculations down to the level of QED contributions. 
\end{abstract}

\maketitle

\section{Introduction}

Transitions of highly charged ions (HCI) of astronomically abundant elements (e.g., oxygen and iron) are essential for the analysis of x-ray observations of warm-hot astrophysical plasmas measured with grating spectrometers onboard the satellites \textit{Chandra} and \textit{XMM-Newton}. The observation of x-ray absorption lines from inner-shell transitions provides diagnostics of the physical conditions of these ionized plasmas, especially their outflow velocities, by observing the wavelength Doppler shift.  Thus, accurate determination of outflow velocities requires both a sufficient spectral resolution and benchmarked reference wavelengths \cite{Beiersdorfer2003}. This diagnostic is often applied with H-like, He-like, and Li-like oxygen ({O}\,\textsc{VIII-VI}) absorption lines to many astrophysical plasmas, revealing for example the inhomogeneous distributions of metals in the hot intra-cluster medium (ICM) \cite{Biffi2018}, the discovery of a large amount of baryonic matter in the local warm-hot intergalactic medium (WHIM) \cite{Nicastro2018}, and the velocity distributions of ionized winds in active galactic nuclei (AGN) \cite{Pinto2018}.  Additionally, the inclusion of lower charge-state lines of oxygen, including the Be-like case ({O}\,\textsc{V}), allows a better determination of the ionic distribution with temperature \cite{Holczer2010} and transition temperature phase \cite{Nevalainen2017}. However, previous laboratory measurements of the {O}\,\textsc{V}  line investigated here, have revealed blending with a {O}\,\textsc{VI} line at $\lambda = 22.374(8)$\,\AA\,\cite{Gu2005}, thereby hindering the usefulness of this line for accurately determining both the ionic and velocity components of the plasma. 
\begin{figure*}[htb]
\includegraphics[width=1\textwidth]{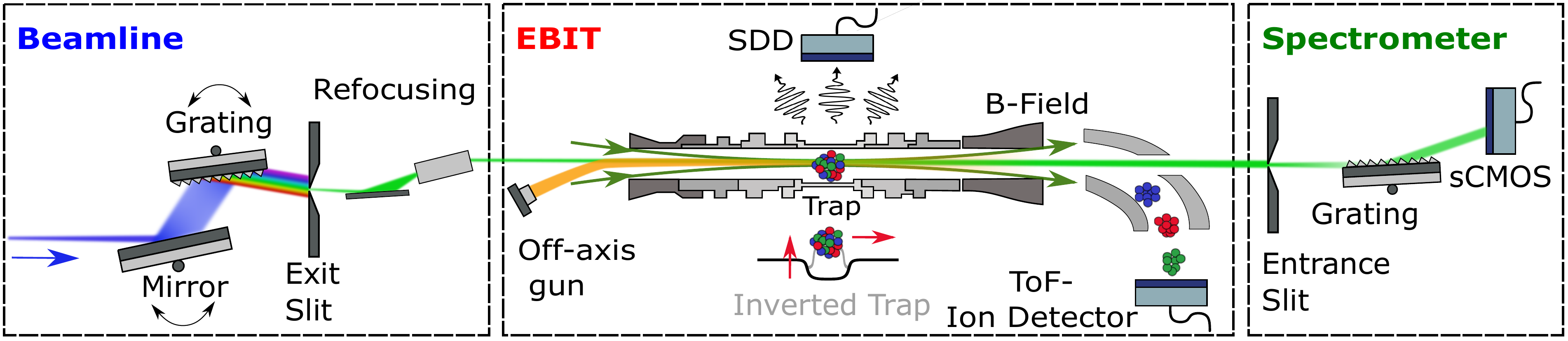}
\caption{Experimental setup. Atoms are ionized and trapped by a magnetically compressed electron beam (orange) emitted by the off-axis electron gun of the portable electron beam ion trap PolarX-EBIT \cite{Micke2018} deployed at the GasPhase soft x-ray undulator beamline of ELETTRA \cite{Prince1998,Blyth1999}. A photon beam from its variable-angle spherical-grating monochromator enters along the EBIT axis and excites the HCI. Fluorescence x rays are recorded by a silicon-drift detector (SDD) mounted side-on. Periodically after a given exposure time, HCI and the generated photoions are ejected and their charge-state distribution determined from the time-of-flight (ToF) data. Downstream, a fixed-angle grating spectrometer records the first-order and second-order diffraction of the photon beam on a sCMOS detector to calibrate its energy independently and unaffected by interpolation errors of the monochromator encoders.
}
\label{Setup}
\end{figure*}
\newline
Furthermore, while hydrogen-like, helium-like and lithium-like ions are in principle well tractable for theory (see, e.~g., calculations in Refs.~\cite{Drake1988,Yerokhin2012,Yerokhin2017,Yerokhin2019,Zaytsev2020}), ions with more bound electrons and semi-filled shells such as the here investigated Be-like oxygen still require numerical approximations\cite{Chen1985,Chen1987,Pradhan2003,Garcia2005,Zhang2012}. While excitation of inner shells such as the $1s\,2s^2\,2p$ studied here are ubiquitous in nature, they involve electronic correlation, QED contributions and couplings to the autoionization continuum that are very difficult to treat theoretically. Therefore, laboratory benchmarks by different methods are needed \cite{Bruch1987,Gu2005,Schmidt2004,Beiersdorfer2003,McLaughlin2016, Machado2018}. 
\newline
Electron beam ion traps (EBIT) in combination with monochromatic x rays from free-electron laser \cite{Epp2007,Bernitt2012} and synchrotron facilities allow for the study of resonant scattering and photoionization \cite{Simon2010,Rudolph2013,Steinbruegge2015} of HCI, yielding highly accurate line positions \cite{Leutenegger2020,Shah2024wave,Togawa2024}, as well as oscillator strengths, charge-exchange cross sections \cite{Shah2016}, and natural linewidths \cite{Kuehn2022,Shah2024gamma}.
\newline
For such experiments, EBITs were equipped with x-ray detectors and a time-of-flight (ToF) ion spectrometer. We additionally introduce a technique to directly measure the photon beam energy using an additional grating spectrometer. In this way, we correct for systematic uncertainties that affect monochromator-based measurements, and reach an outstanding accuracy of 5~parts-per-million (ppm) in the measurements of $K{\alpha}$ transition ($1s^2 2s^2 \rightarrow 1s 2s^2 2p_{3/2}$) in Be-like oxygen O$^{4+}$ ions, as well as allow us to resolve soft x-ray isotopic shifts. Both these quantities serve as stringent tests for calculations, including contributions from quantum electrodynamics (QED). Previous measurements of the same line in middle-$Z$ elements using crystal spectroscopy \cite{Machado2018, Schlesser2013} have also achieved ppm-level accuracy. However, these techniques cannot be applied to energies below 1 keV, where diffraction crystals suffer from reduced reflectivity and resolving power. The present setup overcomes these limitations by operating effectively in the soft x-ray regime.
\section{Measurements and Analysis}%
Our setup (Fig.\ref{Setup}) has three major components: an EBIT for producing and trapping HCI, detecting their fluorescence and determining their charge state; the monochromatic soft x-ray beam from the GasPhase beamline \cite{Prince1998,Blyth1999} focused onto the ion ensemble, and a downstream grating spectrometer that independently measures the photon energy, thereby reducing systematic energy uncertainties stemming from angular encoders of the monochromator -- a novel feature of this setup. PolarX-EBIT \cite{Micke2018} produces HCI by electron-impact ionization. It uses a monoenergetic beam emitted from an off-axis cathode that is magnetically compressed to below \SI{100}{\micro\metre} diameter using an \SI{870}{\milli\tesla} field generated by permanent magnets. In the present experiment we set a low current on the order of \SI{1}{\milli\ampere} to reduce the temperature and Doppler width of the radiation emitted by the trapped ions. This beam radially traps the HCI by virtue of its space charge potential, while the trap drift tubes confine them in the axial direction, resulting in a \SI{22}{\milli\metre} long HCI cloud with a radius of \SI{0.1}{\milli\metre} \cite{Micke2018}. There, the charge-state distribution is governed by competing electron-impact ionization and recombination processes, the residual gas density, and ion evaporative cooling from the trap \cite{Penetrante1991a,Penetrante1991b}. Target species other than residual-gas atoms are injected as tenuous atomic or molecular beams.
\newline
We employ two complementary diagnostic techniques. For the fluorescence, a silicon-drift detector (SDD) with an energy resolution of approximately \SI{100}{\electronvolt} (FWHM) is mounted side on, and shielded from stray light by a \SI{500}{\nano\metre} aluminum foil. For the charge state distribution of the HCI, we employ a time-of-flight (ToF) technique. A high voltage pulse with a rise time of few nanoseconds ejects the complete ion inventory from the trap every second. The ions are accelerated from the positive bias potential of the drift tubes and guided through a Sikler lens \cite{Sikler2011} and an electrostatic $90^\circ$ bender \cite{Kreckel2010} to a channeltron. An oscilloscope records its signal, which sequentially shows the arrival times of different ion species scaling with $\sqrt{m/q}$, where $q$ and $m$ are the ion charge and mass, respectively. The peak amplitude is proportional to the ion number. Small potential differences within the extended ion ensemble, velocity dispersion due to the translational temperature of the trapped ions \cite{Penetrante1991a,Penetrante1991b} and other effects broaden the ToF peaks, the positions of which we calibrate with well-identified ion species as references.
\begin{figure*}[tbh]
\includegraphics[width=0.9\textwidth]{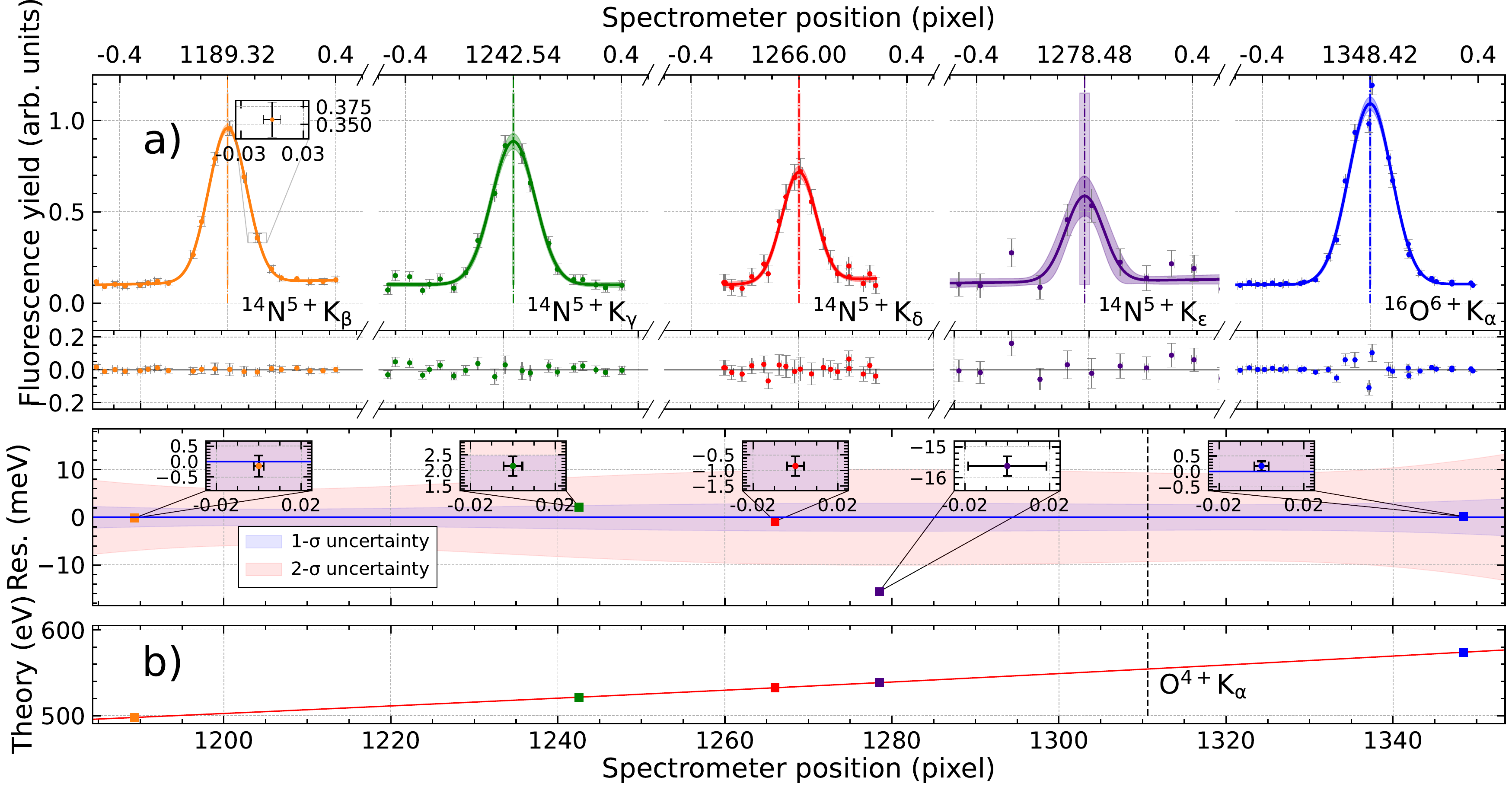}
\caption{
Calibration of grating spectrometer using $\mathrm{N^{5+}}$ (\(K{\beta} - K{\epsilon}\)) and the \(K{\alpha}\) line of 
$\mathrm{O^{6+}}$. 
\textbf{a)}: Fluorescence peaks recorded with the silicon-drift detector with Voigt fits. 
\textbf{b)}: Fitted centroids calibrated using Eq.~\eqref{calibration_function} with experimental and theoretical values from End Matter, Table ~\ref{CalibrationTable}. 
Blue and red shading indicates the 1-$\sigma$ and 2-$\sigma$ confidence bands, respectively.
A vertical dashed line marks the blend of K${\alpha}$ for $\mathrm{^{16}O^{4+}}$ and $\mathrm{^{18}O^{4+}}$.}
\label{Calibration}
\end{figure*}
The experiment was conducted at the GasPhase beamline \cite{Melpignano1995,Prince1998,Blyth1999} at the ELETTRA synchrotron in Trieste, Italy. Linearly polarized photon beams were produced using an undulator. A monochromator, comprising a variable-angle-spherical grating and a plane mirror, was employed to select the desired photon energy. The photon beam was then collimated by a 20\,$\mu$m exit slit and focused onto the ion cloud within an EBIT using a spherical mirror for horizontal and an elliptical mirror for vertical focusing. This setup achieved a resolving power of up to $E/\Delta E \approx 10^4$~\cite{Blyth1999} over the energy range from \SI{360}{\eV} to \SI{900}{\eV}.
\newline
The angles of grating and mirror are controlled by stepper motors and monitored with two Heidenhain linear optical encoders. Their limited absolute accuracy causes energy-dependent interpolation errors that were originally characterized at this beamline with a laser interferometer \cite{Krempasky1997}. 
They result in periodic deviations between nominal and actual photon energies. The same deleterious effects were observed at other facilities in earlier studies \cite{Follath2010, Krempask2011,Kuehn2022}, and were only recently remedied at PETRA III using photoelectron spectrometry \cite{Togawa2024,Shah2024wave}. 
\newline
We here introduce a straightforward, equally effective alternative approach not affected by photoelectron spectrometer noise to solve this problem. For this, we install a grazing-incidence grating spectrometer \SI{2.3}{\m} downstream of PolarX \cite{Poletto2014,Frassetto2011}. It consists of an entrance slit, a spherical variable-line-spacing grating (2400\,grooves/mm) at a fixed angle, and a scientific complementary metal-oxide-semiconductor (sCMOS) detector with $2048\times 2048$ pixels of \SI{11}{\micro\meter} size fully covering the almost-flat focal plane. 
There, line positions depend on the true photon-beam energy independently of beamline settings and monochromator-encoder errors.
The entrance slit solely limits the photon flux to prevent sCMOS saturation and remains fixed throughout the experiment. 
We calibrate the spectrometer dispersion by assigning the energies of theoretically well-known HCI transitions to the corresponding line positions, analogously to the approach used with a photoelectron spectrometer in~\cite{Togawa2024,Shah2024wave}. 
At each step of a photon energy scan, under fixed EBIT parameters, thousand of background-subtracted sCMOS images are averaged, projected, and the resulting spectra are fitted with Voigt profiles. The actual photon energy $E=\frac{hc}{\lambda}$ is then determined from the centroid position $x$ of the first or second-order diffraction peak using the calibration curve,
\begin{equation}
E={ -\frac{h c} {m g\{\sin\left[\frac{\pi}{2}-\arctan(\frac{a-x}{d})\right]-\sin\theta_{\mathrm{in}}\} }},
\label{calibration_function}
\end{equation} 
derived from the grating equation. Here, $\theta_{\text{in}}$ is the angle of incidence relative to the grating normal; $g$ the grating constant, $m$ the diffraction order, $d$ the distance between grating and sCMOS, $a$ the offset of the sCMOS position scale, and the term $\frac{\pi}{2}$ corrects for the exit angle being referenced to the horizontal.
In spite of the lines showing a full-width-at-half-maximum (FWHM) on the sCMOS of $ >\SI{1}{\eV}$, the high counting statistics ($>\SI{1e5}{} \text{counts/s}$) yield uncertainties as low as \SI{5}{\milli \eV} on each step of 60\,s exposure. 

\begin{figure*}[t]
\includegraphics[width=0.9\textwidth]{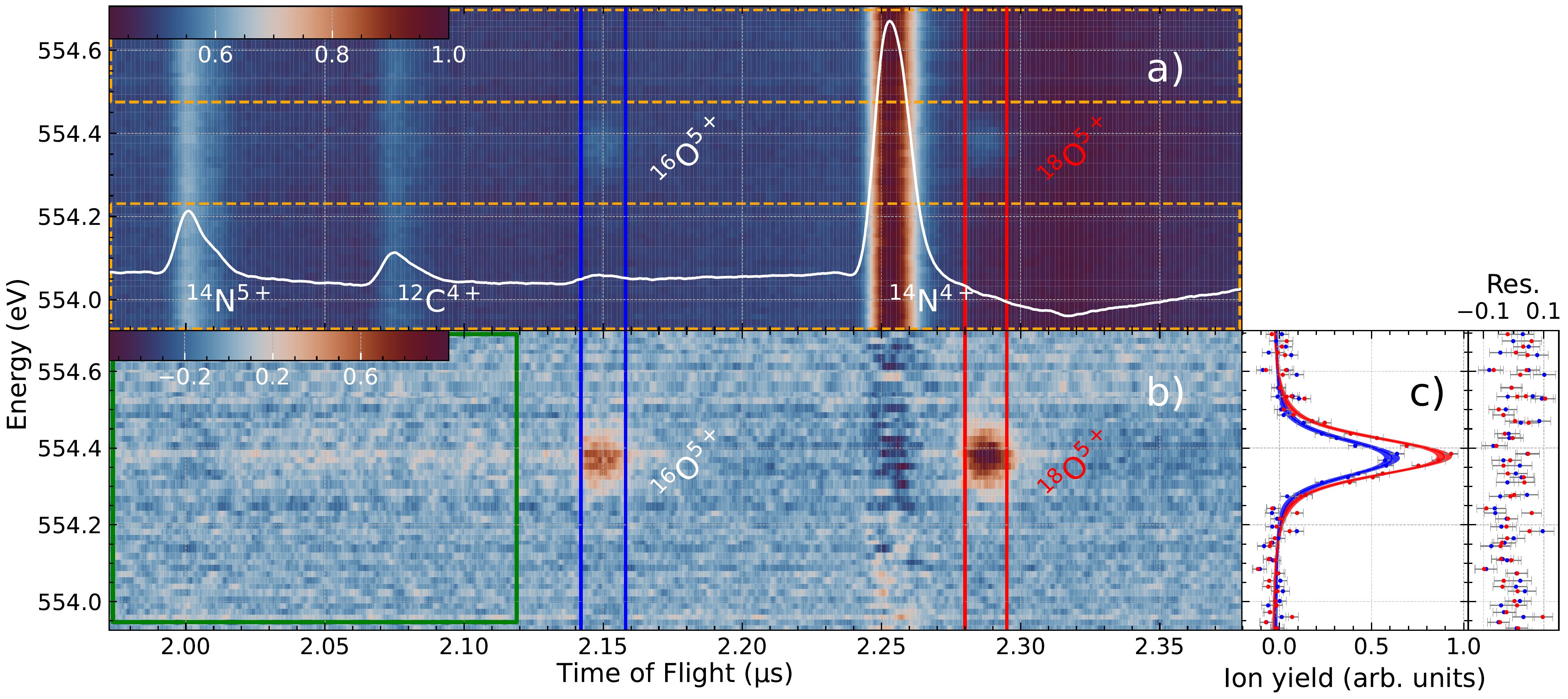}
\caption{One of eight scans recording photoionization of Be-like $^{16}$O$^{4+}$ and $^{18}$O$^{4+}$ ions by the $K{\alpha}$ transition.
(a) Time-of-flight (ToF) data vs.~photon energy, showing resonant and non-resonant features. White curve: Total projected ToF signal.
(b) Subtracting the mean of the projection from non-resonant regions, enclosed in orange-dashed rectangles in panel a), clearly reveals the photoionization resonances.
(c) Photoion yields vs.~calibrated photon energy, projected from the regions of interest for $^{16}$O$^{5+}$ (blue) and $^{18}$O$^{5+}$ (red) in panels (a) and (b). For each data point, we include a relative uncertainty of 0.5\% estimated from the standard deviation of the background counts contained within the green frame in panel (b).
}
\label{IonData_OverView}
\end{figure*}

\begin{figure}[t]
\includegraphics[width=1\columnwidth]{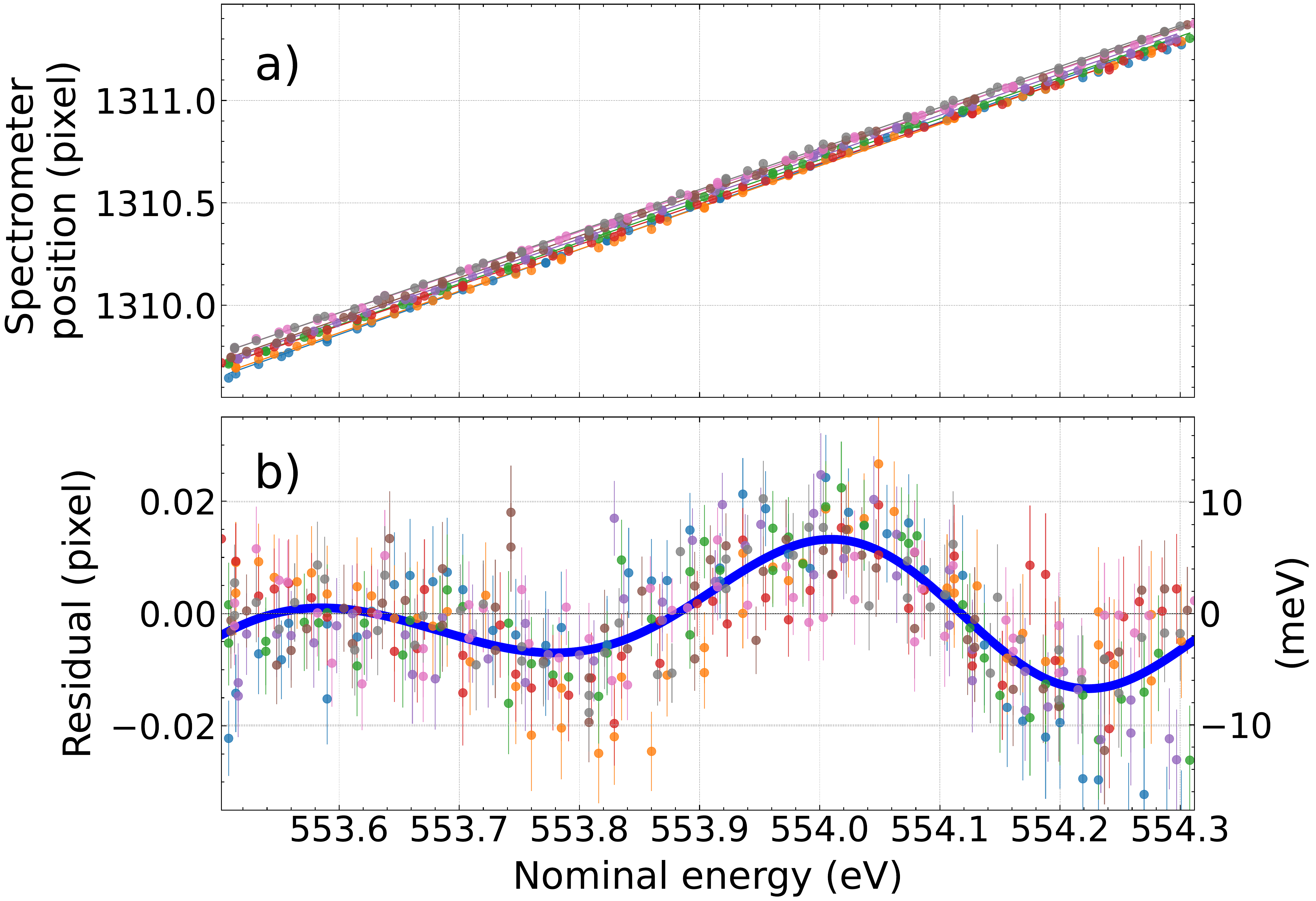}
\caption{(a) Fitted spectrometer line-centroid positions as a function of the nominal photon energy for eight different scans, obtained using the inverse of Eq.~\eqref{calibration_function}. 
(b) Residuals corresponding to each scan. The blue line serves as a guide to the eye and illustrates the reproducible angular encoder error of the monochromator, which is corrected using the spectrometer data. The right y-axis shows the corresponding correction in the actual energy scale.
}

\label{Spectrometer_Res}
\end{figure}

For the energy calibration of the grating spectrometer (Fig.~\ref{Calibration}) around the $\mbox{O}^{4+}$ transition at \SI{554}{\eV}, we use five K-shell transitions: the Rydberg series of He-like nitrogen ($K{\beta}$ to $K{\epsilon}$, \SI{498}{\eV}–\SI{538}{\eV})~\cite{Yerokhin2019} and the $K{\alpha}$ transition of He-like oxygen at $\sim$\SI{574}{\eV}~\cite{Yerokhin2022}. 
They have theoretical uncertainties well below \SI{1}{\milli\eV}, and strongly fluoresce upon resonant photoexcitation. 
Figure~\ref{Calibration}a) shows the total SDD counts within regions of interest containing each of the expected resonances, the statistical uncertainties, and the spectrometer peak position uncertainty at each photon energy. 
The fluorescence yield as a function of spectrometer position is analyzed using orthogonal distance regression (ODR) ~\cite{boggs1990odr}, which accounts for both of those uncertainties, employing a Gaussian function as the fitting model to extract the resonance energies.
In Fig.~\ref{Calibration}b, the peak positions on the sCMOS combined with the theoretical transition energies (see Table~\ref{CalibrationTable}) are used to calibrate the photon-energy scale of the beamline with Eq.~\eqref{calibration_function}, where $m$ and $g$ are fixed parameters.

\begin{table}[ht]
\centering
\caption{Calibration lines with theoretical energies and positions on the spectrometer from Fig.~\ref{Calibration}a.}
\label{CalibrationTable}
\sisetup{
  table-format=4.4,
  table-number-alignment=center
}
\begin{tabular}{
l
l
l
l
S[table-format=4.4(4)]
c
}
\toprule
\textbf{Ion} & \textbf{Label} & \textbf{Lower} & \textbf{Upper} & \textbf{This work (Pixel)} & \textbf{Theor. energy (eV)} \\
\midrule
$N_\mathrm{VI}$ & $K\beta$     & $1s^2$   & $1s\,3p_{3/2}$ & 1189.3228(23) & 497.9194(3) \textsuperscript{a} \\
$N_\mathrm{VI}$ & $K\gamma$    & $1s^2$   & $1s\,4p_{3/2}$ & 1242.5369(43) & 521.5635(3) \textsuperscript{a} \\
$N_\mathrm{VI}$ & $K\delta$    & $1s^2$   & $1s\,5p_{3/2}$ & 1266.0016(40) & 532.5293(3) \textsuperscript{a} \\
$N_\mathrm{VI}$ & $K\epsilon$  & $1s^2$   & $1s\,6p_{3/2}$ & 1278.4838(185)& 538.4925(3) \textsuperscript{a} \\
\midrule
$O_\mathrm{VII}$& w            & $1s^2$   & $1s\,2p_{3/2}$ & 1348.4186(34) & 573.9611(1) \textsuperscript{b} \\
\bottomrule
\end{tabular}
{\begin{flushleft}
\footnotesize
\textsuperscript{a} Ref:~\cite{Yerokhin2019}
\textsuperscript{b} Ref:~\cite{Yerokhin2022}\\
\end{flushleft}}
\end{table}

\begin{figure*}
\centering
\vspace{2mm}
\includegraphics[width=0.9\textwidth]{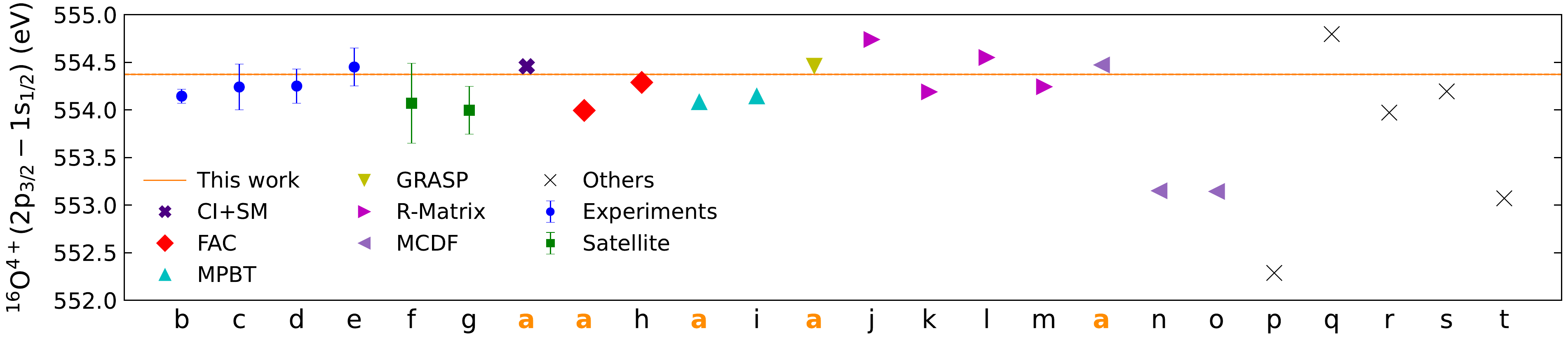}
\caption{
Comparison of the $^{16}$O$^{4+}$ result (orange line) with those of other experiments and observations, as well as with predictions obtained with various methods in this (orange marked) and other works (for details see Table ~\ref{result_table}). The uncertainty of $\pm 0.003\,\mathrm{eV}$ in the present measurement is invisible at this scale.
}
\label{results}
\end{figure*}

\begin{figure}[ht]
\includegraphics[width=1\columnwidth]{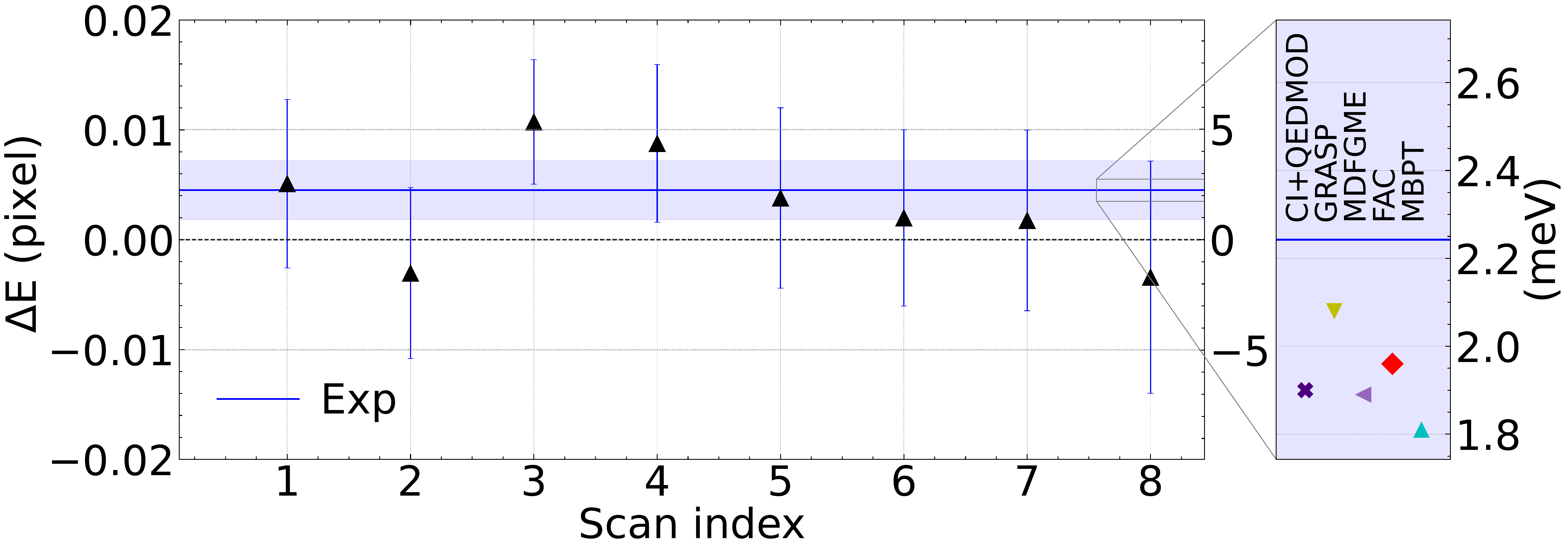}
\caption{
Isotopic shifts $\Delta E$ ($^{18}\mathrm{O}_{K\alpha}^{4+}$-$^{16}\mathrm{O}_{K\alpha}^{4+})$ from eight scans and weighted mean together with theoretical calculations (also in Table ~\ref{result_table}) in the inset. We convert position to photon-energy differences using the slope of the calibration curve [Fig.~\ref{Calibration}b] at the position of the mean of the two resonances.
}
\label{Isotope_Shift}
\end{figure}

\begin{table*}[ht]
\centering
\caption{Experimental and theoretical resonance energies  for the $K_{\alpha}$ transition $1s^2 2s^2~^1\mbox{S}_0 \rightarrow 1s 2s^2 2p_{3/2}~^1\mbox{P}_1 $  in Be-like $^{16}\mathrm{O}^{4+}$ and $^{18}\mathrm{O}^{4+}$ and measured isotope shift (with our results in bold) in comparison with selected experimental and theoretical literature values. All energies are given in eV units, with exception of $\Delta E (^{18}\mathrm{O}^{4+}_{K_\alpha}-^{16}\mathrm{O}^{4+}_{K_\alpha})$ values being in meV units.}
\sisetup{
  table-format=3.3,
  table-number-alignment=center
}
\begin{tabular}{
cccccccccc
}
\toprule
\textbf{Ion} & \textbf{Experiment} & \textbf{Satellite} & \textbf{CI+SM} & \textbf{FAC-CI} & \textbf{FAC-MBPT} & \textbf{GRASP} & \textbf{R-Matrix} & \textbf{MCDF} & \textbf{Others} \\

\midrule
$^{16}$O$^{4+}_{K_\alpha}$
& \textbf{554.372(3)}\textsuperscript{a}  & 554.070(420)\textsuperscript{f}     & \textbf{554.457(9)}\textsuperscript{a}  & \textbf{553.994(250)}\textsuperscript{a}     & \textbf{554.081}\textsuperscript{a}     & \textbf{554.459(52)}\textsuperscript{a}  & 554.739\textsuperscript{j}     & \textbf{554.470(26)}\textsuperscript{a}  & 552.287\textsuperscript{p}     \\

& 554.144(74)\textsuperscript{b} & 553.966\textsuperscript{g}     &              & 554.288\textsuperscript{h}     & 554.144\textsuperscript{i}     &              & 554.189\textsuperscript{k}     & 553.150\textsuperscript{n}      & 554.796\textsuperscript{q}      \\
& 554.24(24)\textsuperscript{c}&             &              &             &             &              & 554.550\textsuperscript{l}     & 553.143\textsuperscript{o}      & 553.971\textsuperscript{r}     \\
& 554.25(18)\textsuperscript{d}&             &              &             &             &              & 554.491\textsuperscript{m}     &              & 554.194\textsuperscript{s}     \\
& 554.45(20)\textsuperscript{e}&             &              &             &             &              & 554.243\textsuperscript{m}     &              & 553.071\textsuperscript{t}     \\
\midrule
$^{18}$O$^{4+}_{K_\alpha}$ 
& \textbf{554.374(3)}\textsuperscript{a}  &             & \textbf{554.459(12)}\textsuperscript{a}  & \textbf{554.996}\textsuperscript{a}     & \textbf{554.083}\textsuperscript{a}     & \textbf{554.461(52)}\textsuperscript{a}  &             & \textbf{554.472(26)}\textsuperscript{a}  &             \\
\midrule
$\Delta E$ (meV) 
& \textbf{2.2(13)}\textsuperscript{a}  &             & \textbf{1.90(2)}\textsuperscript{a}         & \textbf{1.96}\textsuperscript{a}       & \textbf{1.81}\textsuperscript{a}       & \textbf{2.08}\textsuperscript{a}         &             & \textbf{1.89}\textsuperscript{a}         &             \\
\bottomrule
\end{tabular}

\label{result_table}

\vspace{0.5mm}
\begin{tabular}{lll}
\textsuperscript{a} \textbf{This work}                           & \textsuperscript{h} FAC \citet{Liao2013}                    &  \textsuperscript{o} MCDF \citet{Chen1987}    \\ 
\textsuperscript{b} EBIT \citet{Schmidt2004}*                     & \textsuperscript{i} MBPT \citet{Gu2005}                     & \textsuperscript{p} CMR-CI  \citet{Zhang2012}    \\         
\textsuperscript{c} EBIT \citet{Gu2005}                          & \textsuperscript{j} R-Matrix \citet{Pradhan2003}             &  \textsuperscript{q} SCUNC \citet{McLaughlin2016}\\
\textsuperscript{d} IPMB \citet{McLaughlin2016}                  & \textsuperscript{k} R-Matrix \citet{McLaughlin2016}          & \textsuperscript{r} COWAN \citet{Kaastra2004} \\
\textsuperscript{e} Auger spectroscopy \citet{Bruch1987}         & \textsuperscript{l} R-Matrix \citet{Berrington1997}         & \textsuperscript{s} HULLAC \citet{Kaastra2004}\\
\textsuperscript{f} Chandra (Mrk 279) \citet{Kaastra2004}        & \textsuperscript{m} R-Matrix \citet{Garcia2005}              &  \textsuperscript{t} SPCR \citet{Lin2002}\\
\textsuperscript{g} Chandra (NGC 5548) \citet{Schmidt2004}       & \textsuperscript{n} MCDF \citet{Chen1985}            &                                 \\  
*superseded by \citet{Gu2005}                                                
\end{tabular}
\end{table*}
Due to the small isotope shift ($\sim$ \SI{2}{\milli\eV}), the $K{\alpha}$ transitions $1s^2 2s^2 \rightarrow 1s 2s^2 2p_{3/2}$ of $\mathrm{^{16}O^{4+}}$ and $\mathrm{^{18}O^{4+}}$ are unresolved by the SDD, owing to its limited resolving power. 
However, this shift can be resolved using the time-of-flight (ToF) of the Li-like photoions ($^{16}$O$^{5+}$ and $^{18}$O$^{5+}$), which are produced by autoionization following photoexcitation.
To avoid generating Li-like ions through electron-impact ionization, we keep the electron beam energy below the ionization threshold for Be-like ions, which is \SI{113.87}{\eV} \cite{NIST_ASD}.
To determine the resonance energies and their isotopic shifts, we scanned the nominal photon energy over the range from \SI{553.5}{\eV} to \SI{554.3}{\eV}, performing eight measurements in 50 steps of \SI{60}{\s} exposure to photoexcite the $K\alpha$ transition in Be-like oxygen ions. 
A representative scan is shown in Fig.~\ref{IonData_OverView}, displaying the $K{\alpha}$ resonances in the ToF of both $^{16}$O$^{4+}$ and $^{18}$O$^{4+}$ ions. The photon-energy scale was calibrated by converting the fitted spectrometer centroid positions using the calibration function, shown in Fig.~\ref{Calibration}b.
This calibration procedure effectively eliminates photon energy fluctuations arising from angular encoder interpolation errors in the monochromator. 
The impact of these fluctuations can be quantified by fitting, for each scan, the inverse of Eq.~\eqref{calibration_function} to the spectrometer line centroids as a function of the nominal photon energy, as shown in Figure~\ref{Spectrometer_Res}a.
If this correction is omitted, i.e., if the nominal energy is used, the resulting fluctuations introduce a systematic error of approximately \SI{10}{\milli\electronvolt} (see Figure~\ref{Spectrometer_Res}b).
We found analogous periodic errors of up to \SI{70}{\milli \electronvolt} in the photon-energy range {550--1100}~eV in our previous studies at other beamlines~\cite{Kuehn2022,Togawa2024,Shah2024wave} by means of a high-resolution photoelectron spectrometer.
\newline
The centroids of the $1s^2 2s^2 \rightarrow 1s 2s^2 2p_{3/2}$ transition in Be-like $^{16}\mathrm{O}^{4+}$ and $^{18}\mathrm{O}^{4+}$ ions were determined by fitting the spectral projections from all eight scans with a Voigt profile, again using ODR. 
The centroids of each ion, determined in pixel units, were combined into a single value via weighted mean and error, subsequently converted to energy values using the calibration (Fig.~\ref{Calibration}b) function.
We get $^{16}\mathrm{O}_{K\alpha}^{4+}={554.372(3)(1)(3)}$\,{eV} and $^{18}\mathrm{O}_{K\alpha}^{4+}={554.374(3)(1)(3)}$\,{eV}, see Table~\ref{result_table}. The total uncertainty (last parenthesis) has two components: a statistical error (first parenthesis) from the centroids of the Voigt fits, and a systematic (second parenthesis) one derived from the 1-$\sigma$ residuals of the calibration fit at the respective resonance position, as indicated by the black dashed line in Fig.~\ref{Calibration}b).
Our final uncertainty of 5 parts-per-million (ppm) has been only rarely achieved \cite{Togawa2024,Shah2024wave} for many-electron ions in the soft x-ray range.
\newline
To determine the isotope shift, the spectrometer positions differences of the two resonances from each scan were averaged, as shown in Fig.~\ref{Isotope_Shift}, with the uncertainty of each difference taken from the quadratic sum of the centroid uncertainties. 
The resulting shift in spectrometer position with its statistical uncertainty was converted into energy units by multiplying with the slope of the calibration curve at the average centroid of the two transitions, also indicated by a black dashed line in Fig.~\ref{Calibration}b).
No additional systematic uncertainty from the calibration was included, as the estimated uncertainty of the slope, $\SI{10}{\milli \eV}/\SI{150}{pixel}, < \SI{7e-5}{ \eV / pixel}$, is negligible for the small shift. 
This yields a final isotope shift of $\Delta E = 2.2 \pm \SI{1.3}{\milli \eV}$. Its uncertainty can be reduced by increasing the number of scans or the spectral resolution of the spectrometer.

\section{Theory}

We calculate the theoretical energies of the $1s^2 2s^2 ~^1\mbox{S}_0$ and $1s 2s^2 2p ~^1\mbox{P}_1$ levels of both $^{16}$O and $^{18}$O isotopes using various codes and compare them with the present experiment. 
\newline
We first calculate these energies using the configuration interaction (CI) and second-order many-body perturbation theory (MBPT) methods implemented in the Flexible Atomic Code (FAC) package~\cite{Gu2008, Gu2006}.  Calculations were done with principal quantum number up to $n=10$, taking into account frequency-dependent generalized Breit interactions~\cite{Breit1929} in both the CI expansion and the MBPT corrections. QED terms such as self-energy (SE) and vacuum polarization (VP) were incorporated in FAC with an approximate treatment of QED effects by the model operator approach, included in the QEDMOD module \cite{Shabaev2015}.
\newline
We also used the General Relativistic Atomic Structure Package (GRASP) \cite{FroeseFischer2019}, and the multiconfiguration Dirac-Fock General Matrix Elements (MDFGME) \cite{Desclaux1975, Indelicato1990} codes based on multi-configuration Dirac-Fock (MCDF) method.
To avoid numerical instabilities, MCDF calculations are carried out sequentially, starting with dominant configurations $1s^2 2s^2$ and $1s 2s^2 2p$,  with further configuration-state-functions of decreasing impact to the atomic wavefunction being added iteratively. Only single and double excitations of core and valence orbitals are considered within this configuration space.  In each iteration, the single and double excitations are incremented by a single integer in the maximum principal quantum number. For each iteration, the starting orbitals are optimized using the MCDF method and then passed to the subsequent iteration. The Breit interaction and VP are incorporated into the Dirac-Fock equations for the MDFGME case. In contrast, for the GRASP case, these effects are included in a final CI step. Additionally,  SE is evaluated within the operator model approach \cite{Shabaev2015} in MDFGME code, while in GRASP this term is calculated within the Welton approximation \cite{Lowe2013}. To ensure convergence, an intermediate step of considering only double (Brillouin) excitations before including single excitations was employed for the MDFGME case. Due to numerical issues, the calculation is constrained to $n_{\mbox{max}}=7$ and $n_{\mbox{max}}=5$ for the GRASP and  MDFGME cases, respectively.  Unrestricted core-valence excitations were considered for the $^1\mbox{S}_0$ state in both codes, while for the $^1\mbox{P}_1$ state, core-valence excitations were excluded in the GRASP calculation, and limited to $n=3$ in the MDFGME calculation. Nuclear recoil for the isotope shift was calculated by the first-order perturbation formulas for normal-mass-shift (NMS) and specific-mass-shift (SMS) terms \cite{Li2012} with nuclear parameters for $^{16}$O and $^{18}$O from the IAEA LiveChart of Nuclides database \cite{Stone2005, Wang2021, Huang2021}.  At the last iteration, the isotope shift changes by less than 0.5\% in both methods. Table~\ref{tab:grasp_results} lists the energies for each iteration along with energy terms contributing to the total energy.  We estimate the uncertainty of our  GRASP and  MDFGME calculations as the energy difference between the last two iterations. Note that this uncertainty only affects the Coulomb and Breit energies and not the NMS and SMS energies to the quoted precision.

\begin{table*}[htb]
\centering
\caption{Comparison of energy levels, convergence with configuration space and binding energy contributions from Coulomb and Breit interaction, self-energy (SE), vacuum polarization (VP), normal mass shift (NMS), and specific mass shift (SMS) as calculated from GRASP and  MDFGME and for $1s^2 2s^2~^1\mbox{S}_0 \rightarrow 1s 2s^2 2p_{3/2}~^1\mbox{P}_1 $ in Be-like $^{16}\mathrm{O}^{4+}$ and $^{18}\mathrm{O}^{4+}$ (in eV).}
\label{tab:grasp_results}
\sisetup{
  table-format=-4.5, 
  table-number-alignment=center
}
\begin{tabular}{
l
  S[table-format=-4.6] @{\hspace{1em}}
  S[table-format=-4.6] @{\hspace{1em}}
  S[table-format=-4.6] @{\hspace{1em}}
  S[table-format=-4.6] @{\hspace{1em}}
  S[table-format=3.6]  @{\hspace{1em}}
  S[table-format=3.6]
}
\toprule
& \multicolumn{2}{c}{\textbf{$^1\text{S}_0$}} & \multicolumn{2}{c}{\textbf{$^1\text{P}_1$}} & \multicolumn{2}{c}{\textbf{$^1\text{P}_1 - {}^1\text{S}_0$}} \\
\cmidrule(lr){2-3} \cmidrule(lr){4-5} \cmidrule(lr){6-7}
& {GRASP} & {MDFGME} & {GRASP} & {MDFGME} & {GRASP} & {MDFGME} \\
\midrule
	$n$=2	&	-1856.7312	&	-1861.2112	&	-1304.3723	&	-1306.6472	&	552.3589	&	554.5641	\\
	$n$=3	&	-1861.3511	&	-1862.2459	&	-1307.8155	&	-1307.8488	&	553.5356	&	554.3971	\\
	$n$=4	&	-1862.0416	&	-1862.4648	&	-1307.9761	&	-1308.0213	&	554.0654	&	554.4435	\\
	$n$=5	&	-1862.3264	&	-1862.5662	&	-1308.0336	&	-1308.0961	&	554.2929	&	554.4701	\\
	$n$=6	&	-1862.4622	&		 		&	-1308.0551	&				&	554.4071	&		\\
	$n$=7	&	-1862.5227	&				&	-1308.0638	&				&	554.4588	&		\\
	\midrule																																					
	E(Coulomb)	&	-1862.8962	&	-1862.9479	&	-1308.1959	&	-1308.2341	&	554.7003	&	554.7138	\\
	E(Breit)	&	0.1657	&	0.1645	&	0.0092	&	0.0115	&	-0.1566	&	-0.1529	\\ 
	E(SE)	&	0.1497	&	0.1594	&	0.0805	&	0.0828	&	-0.0692	&	-0.0766	\\ 
	E(VP)	&	-0.0069	&	-0.0068	&	-0.0039	&	-0.0040	&	0.0030	&	0.0028	\\
	\midrule													
E(NMS)	$^{16}$O	&	0.0641	&	0.0639	&	0.0437	&	0.0449	&	-0.0204	&	-0.0190	\\
E(SMS)	$^{16}$O	&	0.0008	&	0.0008	&	0.0026	&	0.0027	&	0.0017	&	0.0019	\\
{\textbf{E(Total)	$^{16}$O}	}&	-1862.5227	&	-1862.5662	&	-1308.0638	&	-1308.0961	&	554.4588	&	554.4701	\\
\midrule														
E(NMS)	$^{18}$O	&	0.0570	&	0.0567	&	0.0388	&	0.0399	&	-0.0182	&	-0.0169	\\
E(SMS)	$^{18}$O	&	0.0007	&	0.0007	&	0.0023	&	0.0024	&	0.0016	&	0.0017	\\
{\textbf{E(Total)	$^{18}$O}}	&	-1862.5299	&	-1862.5734	&	-1308.0690	&	-1308.1015	&	554.4609	&	554.4720	\\
\midrule														
E(NMS)	$^{18}$O - $^{16}$O	&	-0.0071	&	-0.0071	&	-0.0049	&	-0.0050	&	0.00228	&	0.00211	\\
E(SMS)	$^{18}$O - $^{16}$O	&	-0.0001	&	-0.0001	&	-0.0003	&	-0.0003	&	-0.00019	&	-0.00022	\\
{\textbf{E(Total)	$^{18}$O - $^{16}$O}}	&	-0.0072	&	-0.0072 &	-0.0051	&	-0.0053	&	0.00208	&	0.00189		\\
\bottomrule
\end{tabular}
\end{table*}
%
Calculations were also performed with the relativistic CI with Stabilization Method (CI+SM) \cite{mandelshtam:93}, having the QED treatment based on the model operator approach \cite{Shabaev2015}. The CI method, as described in Ref.~\cite{Yerokhin2017},  solves the no-pair Hamiltonian with a frozen-core Dirac-Fock $B$-spline basis with Coulomb plus Breit interactions, as well as with relativistic nuclear recoil operators. 
In our implementation of the SM method, a continuous parameter $\gamma$ (defined in Sec.~II of Ref.~\cite{Yerokhin2017}) is introduced in the construction of the discretized one-electron Dirac spectrum. By varying $\gamma$, one effectively shifts the positions of the continuum states in the discretized pseudospectrum. The stabilization method analyzes the dependence of the CI energy eigenvalue $E$ of the $1s2s^22p$ resonance on the parameter $\gamma$, by constructing the so-called stabilization diagram $E(\gamma)$. The resonance position is then identified at the value $\gamma = \gamma_0$ that minimizes the modulus of the derivative $|E'(\gamma)|$. By repeating this procedure for increasing numbers $n$ of $B$-splines in the basis set, we obtain a sequence of energy values $E_n(\gamma_{0,n})$, from which the uncertainty is estimated based on the spread of the results. The CI+SM calculations are tabulated in Table~\ref{TableCI+SM}.

\begin{table*}[htb]
\centering
\caption{Comparison of energy levels and contributions from field shift (FS) and mass shift (MS, combining both NMS and SMS) calculated with CI+SM  for $1s^2 2s^2~^1\mbox{S}_0 \rightarrow 1s 2s^2 2p_{3/2}~^1\mbox{P}_1$ in Be-like $^{16}\mathrm{O}^{4+}$ and $^{18}\mathrm{O}^{4+}$ (in eV).}
\label{tab:CI_SM}
\sisetup{
  table-format=-4.5,
  table-number-alignment=center
}
\begin{tabular}{
l
  S[table-format=-4.7(2)] @{\hspace{5em}}
  S[table-format=-4.7(2)] @{\hspace{5em}}
  S[table-format=3.7(3)] @{\hspace{3em}}
}
\toprule
& {\textbf{$^1\text{S}_0$}} & {\textbf{$^1\text{P}_1$}} & {\textbf{$^1\text{P}_1 - {}^1\text{S}_0$}} \\
\midrule																									
E(Coulomb)	&	-1863.1038(28)	&	-1308.4254(44)	&	554.6783(52)	\\
E(Breit)	&	0.1591(9)	&	0.0091(2)	&	-0.1500(9)	\\ 
E(Recoil)	&	0.0647	&	0.0475	&	-0.0172\\ 
E(QED)	&	0.1323(66)	&	0.0782(39) &	-0.0541(77)	\\
\midrule													
\textbf{E(Total) $^{16}$O}	&	-1862.7477(72)	&	-1308.2906(59)	&	554.4571(93)	\\
\midrule																									
E(FS) $^{18}$O - $^{16}$O	&	0.00002&	0.00001&	-0.00001\\
E(MS) $^{18}$O - $^{16}$O	&	-0.00721(7)	&	-0.00530(5)&	0.00191(2)	\\
\textbf{E(Total) $^{18}$O - $^{16}$O}	&	-0.00718(7) &	-0.00528(5)& 0.00190(2)\\
\bottomrule
\end{tabular}
\label{TableCI+SM}
\end{table*}

\section{Comparison}
We compare in Fig.~\ref{results} our experimental and theoretical values with prior works, reaching an accuracy of 5 parts-per-million, at least twenty times better than earlier experiments in EBITs \cite{Schmidt2004, Gu2005}, ion–photon merged-beam setups \cite{McLaughlin2016}, by Auger spectroscopy \cite{Bruch1987}, as well as the astrophysical observations of Mrk 279 \cite{Kaastra2004} and NGC 5548 \cite{Schmidt2004} by \textit{Chandra}. Overall, all measurements agree with our value, except for a slight deviation from that of Ref.~\cite{Schmidt2004}, which was superseded by \cite{Gu2005} having larger uncertainties. This is due to a blend with Li-like $1s2s2p_{3/2,1/2}\rightarrow 1s^2 2s$ transitions at $\lambda = 22.374(8)$\,\AA\ \cite{Gu2005}, not accounted for previously \cite{Schmidt2004}. 
Various theoretical predictions, including ours, scatter around our experimental
result, differences being larger than the estimated theoretical error bars. 
Those from FAC, CI+SM, MBPT, GRASP, MDFGME, and R-matrix methods cluster around our result, while the cited MCDF \cite{Chen1985,Chen1987}, CMR-CI \cite{Zhang2012}, and saddle-point calculations (SPCR) \cite{Lin2002} show larger discrepancies, suggesting limitations in their approximations. 
The possible reason for the observed discrepancies could be
an incomplete description of the interaction of the autoionizing $1s2s^22p$ resonance with the positive-energy continuum in the calculations. 
A straightforward application of standard atomic structure methods to such a resonance can produce large (and essentially random) energy shifts due to this continuum interaction. This effect likely accounts for the significant spread among the various predictions. Even approaches that account for the continuum interaction of autoionizing states using the stabilization method, such as CI+SM, can miss part of the continuum effects~\cite{zaytsev:19,zaytsev:23}. This underscores the need for \textit{ab initio} calculations of the $1s2s^22p$ resonance using the complex rotation method, as has been demonstrated for helium-like ions~\cite{zaytsev:19,zaytsev:23}.
\newline
As for the isotopic shift, since no experimental data exist for the $K\alpha$ energy of $^{18}\mathrm{O}^{4+}$ measured in this work, we can only compare the result for $^{18}\mathrm{O}^{4+}$ and the isotopic shift with our own calculations. 
As shown in Fig.~\ref{Isotope_Shift}, our predictions are systematically offset from our experimental result by approximately \SI{0.3}{\milli \eV}, but well within its error bar.

\section{Summary and Conclusions}
Using resonant photoexcitation at the ELETTRA synchrotron combined with an electron beam ion trap, we measured the inner-shell transition energy
\(1s^2 2s^2\,{}^1\!S_0 \rightarrow 1s\,2s^2 2p_{3/2}\,{}^1\!P_1\)
of \(^{16}\mathrm{O}^{4+}\) (\(K\alpha\)) to be 554.372(3)\,\si{\electronvolt}.
We compared this value with our most advanced electronic-structure calculations based on the CI+SM approach and found a disagreement of approximately 85\,meV, whereas previously published theoretical values show a scatter exceeding \(\pm 250\)\,meV.
In addition, we measured the same transition in the \(^{18}\mathrm{O}\) isotope and observed a small isotope shift of \(2.2 \pm 1.3\)\,meV in the soft x-ray regime with our time-of-flight setup.
\newline
In general, our measurement method improves the accuracy of energy determinations for autoionizing transitions by more than one order of magnitude and reveals previously unresolvable isotopic shifts at soft x-ray energies. 
Here, it provides for $^{16}\mathrm{O}^{4+}$ high-accuracy data on an essential highly charged ion needed for x-ray astrophysics, and potentially for fusion-plasma applications. Moreover, by suppressing systematic uncertainties in the energy calibration that have affected most previous studies using monochromators, our method enables more stringent tests of atomic structure theory including QED and nuclear recoil, as shown here in for a, relatively simple, four-electron system. 
\section*{Acknowledgments}
Research was funded by the Max-Planck-Gesellschaft (MPG), Germany. The research leading to these results has received partial funding from the European Union's Horizon 2020 Programme under the AHEAD2020 project (grant agreement n. 871158 and n. 654215). 
We acknowledge Elettra Sincrotrone Trieste for providing access to its synchrotron radiation facilities.
P.A. and F.G. acknowledge support from Funda\c{c}\~{a}o para a Ci\^{e}ncia e Tec\-no\-lo\-gia, Portugal, under contracts No. LA/P/0117/2020 (REAL), FCT/Mobility/1301286862/2024-25 and UI/BD/151000/2021. P.A. expresses his thanks for discussions and feedback from José P.~Marques concerning the use of MDFGME. C.S. acknowledges support from MPG and NASA-JHU Cooperative Agreement.

\bibliographystyle{apsrev4-2}
\bibliography{Bib}

\end{document}